\documentclass[12pt]{article}
\usepackage[margin = 1in]{geometry}
\usepackage[utf8]{inputenc}
\usepackage{fullpage}
\usepackage{microtype}
\usepackage{hyperref}
\usepackage[round]{natbib}
\usepackage{graphicx}
\usepackage[dvipsnames]{xcolor}
\usepackage{algorithm}
\usepackage{multirow}
\usepackage{amsmath, amssymb, amsthm}
\usepackage{setspace}
\usepackage{bm}

\theoremstyle{plain}
\newtheorem{theorem}{Theorem}[section]

\newtheorem{corollary}[theorem]{Corollary}
\theoremstyle{definition}

\newtheorem{assumption}{Assumption}
\newtheorem{example}{Example}
\theoremstyle{remark}

\newcommand{\E}{\mathbb{E}}
\newcommand{\I}{\mathbb{I}}

\newcommand{\cov}{\operatorname{cov}}

\newcommand{\pr}{\mathsf{P}}
\newcommand{\dd}{\mathrm{d}}

\newcommand{\tsp}{\mathsf{T}}
\newcommand{\rN}{\mathcal{N}}

\newcommand{\emin}{\lambda_{\min}}
\newcommand{\R}[1]{\mathbb{R}^{#1}}
\DeclareMathOperator*{\argmin}{arg\,min}

\begin{document}

\title{Concave likelihood-based regression with finite-support response variables}
\author{Karl Oskar Ekvall$^{\star,\dagger}$ \quad Matteo Bottai$^\star$\\
{\tt \normalsize  \quad \quad k.ekvall@ufl.edu \quad \quad\quad matteo.bottai@ki.se} \\
{\normalsize $^\star$Division of Biostatistics, Institute of Environmental Medicine, Karolinska Institutet} \\
{\normalsize $^\dagger$Department of Statistics, University of Florida}}
\date{}

\maketitle

\onehalfspacing

\begin{abstract}
  We propose a unified framework for
  likelihood-based regression modeling when the response variable has finite
  support. Our work is motivated by the fact that, in practice, observed data
  are discrete and bounded. The proposed methods assume a model which includes
  models previously considered for interval-censored variables with log-concave
  distributions as special cases. The resulting log-likelihood is concave, which
  we use to establish asymptotic normality of its maximizer as the number of
  observations $n$ tends to infinity with the number of parameters $d$ fixed,
  and rates of convergence of $L_1$-regularized estimators when the true
  parameter vector is sparse and $d$ and $n$ both tend to infinity with $\log(d)
  / n \to 0$. We consider an inexact proximal Newton algorithm for computing
  estimates and give theoretical guarantees for its convergence. The range of
  possible applications is wide, including but not limited to survival analysis
  in discrete time, the modeling of outcomes on scored surveys and
  questionnaires, and, more generally, interval-censored regression. The
  applicability and usefulness of the proposed methods are illustrated in
  simulations and data examples.
\end{abstract}

\doublespacing

\section{Introduction} \label{sec:intro}

  In practice observed data are discrete and bounded, be it by design, because
  of limited measurement precision, or because the data are stored in finite
  precision, for example as floating point numbers in a computer. However, it is
  common to ignore this and use models assuming continuous distributions, or
  continuous models for short. In general this practice leads to
  misspecification and biased estimators. While the bias
  can be small in some settings, it can be substantial in others, and the
  practice nevertheless persists. In our experience, this is in part due to a
  lack of reliable methods for the correctly specified likelihood and an
  unawareness of the potential pitfalls. To address these issues, we propose
  methods with theoretical and computational guarantees for a flexible class of
  regression models for finite-support (i.e., discrete and bounded) response
  variables. In addition, we illustrate the bias that can result from
  incorrectly applying a continuous model using simulations.

	We consider four data examples, two of which are provided as Supporting
	Information. Each example is in a different setting with different challenges.
	Given the ubiquitous use of continuous models with data with finite support,
	the examples can illustrate but a small fraction of the many potential
	applications for the proposed methods. The first data example (Section
	\ref{sec:plasma}) focuses on the effects of clinical predictors on plasma
	lipoprotein(a) [Lp(a)] levels measured in clinical care. Lp(a) is measured
	with finite precision, has a lower limit of detection, and a natural upper
	bound. Thus, in practice Lp(a) has finite support.

	In the second example, cancer patients are observed repeatedly over the course
	of a study. For each patient, the time to death or distant metastases is
	recorded. Interest can be in univariable modeling of the time-to-event or, as
	is the focus in Section \ref{sec:cancer}, the effects of clinical predictors
	and prediction using gene expressions. Either way, time-to-event
	has finite support: patients do not live forever, and time is measured
	with finite precision. Additionally, in many studies patients can be observed
	only at a few specific time points, leading to the observable time-to-event
	being far from continuous.

	The third example (Web Appendix A.1) has an ordinal response, the total score
	on a depression screening questionnaire, taking values in $\{0, 1, \dots,
	27\}$, and illustrates how the proposed methods can be used in settings where
	there need not exist a latent continuous variable of interest. The fourth
	example (Web Appendix A.2) focuses on discovering genes that predict, or are
	associated with, glucose intolerance. The response is ordinal with three
	levels and the number of predictors is three orders of magnitude larger than
	the number of observations.
  
	Now, regardless of application, any response variable $Y$ with finite support
	$\mathcal{Y}$ can be modeled using the categorical distribution parameterized
	by the category probabilities. When the number of categories, that is, the
	cardinality of $ \mathcal{Y}$, is small relative to the number of
	observations, it may be possible to estimate those probabilities with
	acceptable precision using the corresponding sample proportions. However, when
	the number of categories is large, their probabilities depend on predictors,
	or there is a known relation between the probabilities, then further modeling
	is often needed. We consider a model which handles many practically
	relevant settings and which leads to estimators with theoretical support.
	Specifically, we assume the probability mass function for $Y$ given a
	non-stochastic predictor vector $\bm{x}\in \R{p}$ can be expressed, for
	functions $a$ and $b$ to be specified, as
  \begin{equation} \label{eq:model}
    f_{{\bm{\theta}}}(y\mid \bm{x}) = \int_{a(y, \bm{x}, {\bm{\theta}})}^{b(y,
    \bm{x}, {\bm{\theta}})} r(w) \, \dd w = R\{b(y, \bm{x}, {\bm{\theta}})\} -
    R\{a(y, \bm{x}, {\bm{\theta}})\},
  \end{equation}
  where $r$ is a log-concave Lebesgue-density on $\R{}$, $R$ the corresponding
  cumulative distribution function, and ${\bm{\theta}}$ a parameter vector. We
  will assume $a$ and $b$ are affine in $\bm{\theta}$ for every $(y, \bm{x})$ and
  give further details on the specification in Section \ref{sec:model}.

	Intuitively \eqref{eq:model} can be understood as the mass function for an
	interval-censored latent, continuous random variable $W$ with density $r$. In
	some settings $W$ has a practical interpretation. For example, it is typically
	related to the unobservable continuous time-to-event in settings such as the
	cancer study discussed above. On the other hand, \eqref{eq:model} is also
	useful in many settings where there is no latent variable of practical
	interest. In fact, Example \ref{ex:categorical} establishes that, when there
	are no predictors, any mass function for a categorical random variable can
  be obtained as a special case of \eqref{eq:model}.

  Authors considering models like \eqref{eq:model} include \citet{Burridge1981,
  Burridge1982} who note that, in some cases of interest, the log-likelihood is
  concave. These and some related results are discussed in the review of methods
  for grouped data by \citet{Heitjan1989}. At the time, much of the literature
  was concerned with adjusting methods for continuous data to address bias
  introduced by grouping. By contrast, the focus here is the development of
  methods based on the correct likelihood. Likelihood-based methods for settings
  related to ours include that by \citet{Finkelstein1986}, who proposed a model
  for interval-censored failure time data. \citet{Gentleman.Geyer1994} gave
  statistical and computational guarantees for maximum likelihood estimates
  under interval-censoring of a non-parametric model for survival times, and
  \citet{Huang1996} provided convergence rates for maximum likelihood estimators
  in interval-censored proportional hazards models. More recently,
  \citet{Taraldsen2011} studied the special case of rounded exponential data in
  detail, \citet{Zeng.etal2016} proposed methods for interval-censored survival
  times, \citet{Couso.etal2017} discussed different coarsening processes, and
  \citet{Guillaume.etal2017} proposed robust optimization methods for coarse
  data in an essentially non-parametric setting. \citet{Kowal.Canale2020} also
  proposed a non-parametric method, for integer-valued data, mentioning rounded
  data as a relevant special case. \citet{McGough.etal2021} studied penalized
  regression for censored and truncated, but not interval-censored, data.
  Notably, many of the applications are in survival analysis, which is natural
  given that time is generally measured in discrete units. A thorough treatment
  of survival analysis in discrete time is given by \citet{Tutz.Schmid2016}.
  Here, we consider a unified framework including some models for survival
  analysis as special cases.

  While some special cases of \eqref{eq:model}, for example logistic regression
  (see Section \ref{sec:model}) and cumulative probability models (Example
  \ref{ex:categorical}), have been studied extensively, the general setting has
  not. We give intuitive conditions on the density $r$ and endpoints $a$
  and $b$ which guarantee asymptotic normality of the maximum likelihood
  estimator when the number of observations $n$ grows with the number of
  parameters $d$ fixed. Essentially, an asymptotic rank condition on a model
  matrix and $r$ being continuously differentiable suffices. We also consider
  settings where $d$ tends to infinity with $\log(d) / n \to 0$, and give
  convergence rates for an $L_1$-regularized maximum likelihood estimator under
  a restricted eigenvalue condition on a model matrix and $r$ continuously
  differentiable. Finally, we establish the numerical convergence of an inexact
  proximal Newton algorithm under conditions similar to those ensuring
  statistical convergence.

\section{Model} \label{sec:model}
  Let $\Theta \subseteq \R{d}$ be a convex parameter set and suppose $Y_1,
  \dots, Y_n$ are independent, each having a mass function consistent with Model
  \ref{eq:model}:
  \begin{equation} \label{eq:model_detailed}
    f^i_{\bm{\theta}}(y_i\mid \bm{x}_i) = \int_{a_i(y_i, \bm{x}_i,
    {\bm{\theta}})}^{b_i(y_i, \bm{x}_i, {\bm{\theta}})}r(w)\, \dd w =
    R\{b_i(y_i, \bm{x}_i, {\bm{\theta}})\} - R\{a_i(y_i, \bm{x}_i,
    {\bm{\theta}})\},
  \end{equation}
  where $\bm{x}_i \in \mathcal{X} \subseteq \R{p}$. We will often write
	$f_{\bm{\theta}}(y_i \mid \bm{x})$ instead of $f^i_{\bm{\theta}}(y_i\mid
	\bm{x}_i)$ for brevity. The support $\mathcal{Y}_i = \{y_i :
	f^i_{\bm{\theta}}(y_i\mid \bm{x}) > 0\}$ need not be the same for every $i$,
	but we will assume $\mathcal{Y} = \cup_{i = 1}^\infty \mathcal{Y}_i$ is
	finite. Define $r(\infty) = r(-\infty) = R(-\infty) = 0$ and $R(\infty) = 1$.
  
  We assume, for $\bm{Z}_i = \bm{Z}_i (y_i, \bm{x}_i) \in \R{2 \times d}$ and
	$\bm{m}_i = \bm{m}_i (y_i, \bm{x}_i) \in [-\infty, \infty]^2$ to be defined
	shortly, $[a_i(y_i, \bm{x}_i, \bm{\theta}),  b_i(y_i, \bm{x}_i,
	\bm{\theta})]^\tsp = \bm{Z}_i \bm{\theta} + \bm{m}_i$. When writing
	$\bm{Z}_i$ and $\bm{m}_i$ for brevity, dependence on $(y_i, \bm{x_i})$ is
	implicit. Denote the first and second element of $\bm{m}_i$ by, respectively,
	$m_i^a$ and $m_i^b$. Accordingly, denote the first and second row of
	$\bm{Z}_i$ by, respectively, $\bm{z}_i^a$ and $\bm{z}_i^b$. We assume that if
	$m_i^a(y_i, \bm{x}_i) = -\infty$ for some $y_i \in \mathcal{Y}_i$, then it
	holds for every $\bm{x}_i \in \mathcal{X}$; and, for those $y_i$, we let
	$\bm{z}_i^a(y_i, \bm{x}_i) = \bm{0}$ for every $\bm{x}_i \in \mathcal{X}$.
	Thus, whether $a_i(y_i, \bm{x}_i, \bm{\theta})$ is finite or not depends only
	on $y_i$. Similarly, if $m_i^b(y_i, \bm{x}_i) = \infty$ for some $y_i \in
	\mathcal{Y}_i$, then it holds for every $\bm{x}_i \in \mathcal{X}$; and
	$\bm{z}^b_i(y_i, \bm{x}_i) = \bm{0}$. 

  The following three examples illustrate definitions and connections to
  some common models. Example \ref{ex:categorical} shows that, when there are
  no predictors, any model for a categorical response is a special case of
  \eqref{eq:model}, while Examples
  \ref{ex:int_cens_reg} and \ref{ex:flex_par_surv} include predictors.

  \begin{example}[Cumulative probability models] \label{ex:categorical}
    
    Consider a response $Y$ with $m$ possible values, without loss
    of generality $\mathcal{Y}  =  \{1, \dots, m\}$. A possible version of
    \eqref{eq:model} assumes $f_{\bm{\theta}}(y)$ is defined by
    \begin{align} \label{eq:pmf_cate}
      \begin{aligned}
      & f_{\bm{\theta}}(1) = R(\theta_1), ~~f_{\bm{\theta}}(2) = R(\theta_2) - R(\theta_1), \dots,~~ f_{\bm{\theta}}(m - 1) = R(\theta_{m - 1}) - R(\theta_{m - 2}), \\
      &f_{\bm{\theta}}(m) = 1 - R(\theta_{m - 1}),
      \end{aligned}
    \end{align}
    with parameter set $\Theta = \{\bm{\theta} \in \R{m - 1}: \theta_{j} \geq
    \theta_{j - 1}, j \in \{2, \dots, m - 1\}\}$. In the notation of
    \eqref{eq:model}, without predictors, $a(y, \bm{\theta}) = -\infty$ if $y =
    1$, and $a(y, \bm{\theta}) = \theta_{y - 1}$ otherwise. Similarly, $b(y,
    \bm{\theta}) = \infty$ if $y = m$ and $b(y, \bm{\theta}) = \theta_{y}$
    otherwise. One may also write \eqref{eq:pmf_cate} as $\pr_\theta(Y \leq j) = R(\theta_j)$, $j \in \{1, \dots, m - 1\}$,
    which shows cumulative probability models are a special case of
    \eqref{eq:model}; see e.g. \citet[Section 6.2]{Agresti2019}, who uses a
    different but equivalent parameterization. Because $R$ is continuous it is
    straightforward to show any vector of category probabilities
    $[f_{\bm{\theta}}(1), \dots, f_{\bm{\theta}}(m)]^\tsp$ is attainable as
    $\bm{\theta}$ varies in $\Theta$. Thus, any categorical distribution is a
    special case of \eqref{eq:model}. Lastly we note that, in this example, any
    choice of $R$ gives the same model, or set of distributions,
    $\{f_{\bm{\theta}}: \bm{\theta} \in \Theta\}$. This will in general not be
    the case when there are predictors as, then, $R$ determines how the
    predictors affect the probabilities.
  \end{example}
  
  \begin{example}[Interval-censored regression] \label{ex:int_cens_reg}
    Suppose for some $\sigma > 0$, ${\bm{\beta}} \in \R{p}$, and $W_i$ with
    log-concave Lebesgue-density $r$ on $\R{}$, independently for $i \in \{1,
    \dots, n\}$, \begin{equation} \label{eq:int_cens_reg}
      Y_i^* = \bm{x}_i^\tsp {\bm{\beta}} + \sigma W_i.
    \end{equation}
    Suppose also, for some $k_i\geq 1$ and known cut points $-\infty = t_0^i
    < t_1^i < \cdots < t_{k_i}^i < t^i_{k_i + 1} = \infty$, the observed
    response is
    \[
      Y_i = \begin{cases} y_i^{(0)}, & Y_i^* \in (t^i_0, t^i_1) \\
   \vdots & \\
    y_i^{(k_i)}, & Y_i^* \in [t_{k_i}^i, t_{k_i + 1}^i)
    \end{cases},
    \]
    where the interval labels $y^{(0)}_i, \dots, y^{(k_i)}_i$ are arbitrary.
    Common binary regression models such as probit and logistic regression are
    special cases with, for every $i$, $k_i= 1$, $t^i_1 = 0$, known $\sigma =
    1$, and $W_i$ having standard normal or logistic distribution, respectively.
    More generally, in the parameterization ${\bm{\theta}} =  [\sigma^{-1},
    \sigma^{-1}{\bm{\beta}}^\tsp]^\tsp \in \R{p + 1}$, for $j \in \{0, \dots,
    k\}$,
    \begin{align} \label{eq:pmf_intcens}
      f_{\bm{\theta}}(y_i^{(j)}\mid \bm{x}_i)  =
      R([t_{j + 1}^i, -\bm{x}_i^\tsp] {\bm{\theta}} ) - R([t_{j}^i, -\bm{x}_i^\tsp]
      {\bm{\theta}}),
    \end{align}
    which is consistent with \eqref{eq:model}. In particular, $a_i(y_i,
    \bm{x}_i, {\bm{\theta}}) = -\infty$ if $y_i = y^i_{(0)}$ and, otherwise,
    $a_i(y_i, \bm{x}_i, {\bm{\theta}}) = {\bm{\theta}}^\tsp \bm{z}_i^a = \bm{\theta}^\tsp [t^i_{j}, -\bm{x}_i^\tsp]^\tsp$ and
    $m_i^a = 0$. Similarly, $b_i(y_i, \bm{x}_i, {\bm{\theta}}) =
    \infty$ if $y_i = y_i^{(k)}$ and, otherwise, $b_i(y_i, \bm{x}_i,
    {\bm{\theta}}) = {\bm{\theta}}^\tsp \bm{z}_i^b =
    {\bm{\theta}}^\tsp [t^i_{j + 1}, -\bm{x}_i^\tsp]^\tsp$ and $m^b_i =
    0$.

   Without predictors, \eqref{eq:pmf_intcens} is similar to \eqref{eq:pmf_cate}.
   However, in \eqref{eq:pmf_intcens} without predictors the arguments to $R$
   are determined by the known cut points and one parameter, $\theta_1$,
   while in \eqref{eq:pmf_cate} the cut points are parameters.
  \end{example}

  \begin{example}[Interval-censored flexible parametric survival models] \label{ex:flex_par_surv}
    \citet{Royston.Parmar2002} introduce a class of flexible parametric models
    for survival analysis. One model assumes a survival time $T$ has cumulative
    distribution function $ F(t; \bm{x}, {\bm{\beta}}, {\bm{\gamma}}) = 1 -
    \exp[-\exp\{\mathtt{sp}(\log t; {\bm{\gamma}})-{\bm{\beta}}^\tsp \bm{x}\}],
    $ where $\mathtt{sp}(\log t; {\bm{\gamma}})$ is a spline of $\log (t)$ with
    coefficients ${\bm{\gamma}}$. Any $\mathtt{sp}(\cdot; {\bm{\gamma}})$ which
    is monotone increasing for every ${\bm{\gamma}}$ in the parameter set and
    tends to $\pm \infty$ when its argument does, gives a valid cumulative
    distribution function. The exponential distribution is a special case with
    $\mathtt{sp}(\log t; {\bm{\gamma}}) = \log t$. In practice, what is observed
    is often an interval containing $T$, say
    \[
      Y = \begin{cases} y^{(0)}, & T \in [0, t_1) \\ \vdots & \\ y^{(k)}, & T
      \in [t_{k}, \infty) \end{cases},
    \]
    where $t_0 = 0 < t_1 < \cdots < t_k < \infty$ the
    observation subscript $i$ is suppressed for simplicity. Thus, for example, $
      f_{\bm{\theta}}(y^{(1)}\mid \bm{x}) = F(t_2; \bm{x}, {\bm{\beta}},
      {\bm{\gamma}}) - F(t_1; \bm{x}, {\bm{\beta}}, {\bm{\gamma}})$.
    Using this it is straightforward to show the mass function for $Y$ satisfies
    \eqref{eq:model} with $R(w) = 1 - \exp\{-\exp(w)\}$, ${\bm{\theta}} =
    [{\bm{\gamma}}^\tsp, {\bm{\beta}}^\tsp]^\tsp$, and $\bm{z}^a$ and
    $\bm{z}^b$ defined accordingly.
  \end{example}

  Next we establish concavity of the log-likelihood. The log-likelihood for one
  observation is $\ell^i({\bm{\theta}}; y_i, \bm{x}_i) = \log
  \{f_{\bm{\theta}}(y_i\mid \bm{x}_i)\}$, and $\ell_n({\bm{\theta}}; \bm{Y},
  \bm{X}) = \sum_{i = 1}^n \ell^i({\bm{\theta}}; y_i, \bm{x}_i)$, where $\bm{Y}
  = [Y_1, \dots, Y_n]^\tsp \in \R{n}$ and $\bm{X} = [\bm{x}_1, \dots,
  \bm{x}_n]^\tsp \in \R{n\times p}$. 

  \begin{theorem} \label{thm:concave} The log-likelihood $\ell_n(\cdot;
    \bm{Y}, \bm{X})$ given by model \eqref{eq:model_detailed} is concave on
    $\Theta$. Moreover, if $\sum_{i = 1}^n \bm{Z}_i^\tsp \bm{Z}_i$ is positive
    definite  and $r$ is strictly positive, strictly log-concave, and
    continuously differentiable; then $\ell_n(\cdot; \bm{Y}, \bm{X})$ is
    strictly concave on every open, convex subset of $\Theta$.
  \end{theorem}

  The proof of Theorem \ref{thm:concave} uses classical results on log-concave
  functions due to \citet{Prekopa1973} and is in the Supporting Information
  along with proofs of other formally stated results. A special case of the non-strict concavity
  given by Theorem \ref{thm:concave} is discussed without proof by \citet[p.150]{Burridge1982}. We have not seen the strict
  part, which requires substantially more work, stated or proved before. The
  essential component in its proof is Lemma B.1 (Supporting Information) which
  establishes strict log-concavity of the map $(t_1, t_2) \mapsto \{R(t_2) -
  R(t_1)\}$. The strictness of that log-concavity is critical for our results
  with diverging number of parameters. 

\section{Asymptotic properties}\label{sec:asy}
\subsection{Fixed number of parameters}
  We consider maximum likelihood estimators
  \[
    \widehat{{\bm{\theta}}}_n \in \argmin_{{\bm{\theta}} \in \Theta}
    G_n({\bm{\theta}}; \bm{Y}, \bm{X}),
  \]
  where $G_n({\bm{\theta}}; \bm{Y}, \bm{X}) = -n^{-1}\ell_n({\bm{\theta}};
  \bm{Y}, \bm{X})$. Because $\ell_n(\cdot ; \bm{Y}, \bm{X})$ is concave on the
  convex $\Theta$ (Theorem \ref{thm:concave}),
  $\widehat{{\bm{\theta}}}_n$ is a solution to a stochastic convex optimization
  problem, which is used in the proofs of our main asymptotic results. 
  
  In results and their proofs $c_j \in (0, \infty)$, $j \in \{1, 2, \dots\}$,
  denote generic constants which can change between statements but, in each
  statement, depend on neither of $i$, $n$, $d$, $\bm{Y}$, $\bm{X}$, or
  $\bm{\theta}$. We use $\Vert \cdot \Vert$ for the spectral norm for matrices
  and Euclidean norm for vectors, $\Vert \cdot \Vert_\infty$ for the max-norm
  (maximum absolute element), and $\Vert \cdot\Vert_1$ for the one-norm (sum of
  absolute elements). The true parameter is denoted $\bm{\theta}_*$.

  The following assumption will be used in both the low- and high-dimensional
  settings.

  \begin{assumption} \label{ass:compact}
    For all small enough $\rho > 0$, there is a compact $E \subseteq \{\bm{t}
    \in \R{2}: t_1 < t_2\}$ such that, for every $i \in \mathbb{N} = \{1, 2,
    \dots\}$, $y_i \in \mathcal{Y}_i$, $\bm{x}_i \in \mathcal{X}$, and
    $\bm{\theta} \in \Theta$ with $\Vert \bm{\theta} - \bm{\theta}_*\Vert_1 \leq
    \rho$, it holds that either $\bm{Z}_i \bm{\theta} + \bm{m}_i \in E$ or an
    element of $\bm{m}_i$ is infinite. Moreover, for some $c_1  < \infty$,
    $\Vert \bm{Z}_i\Vert_\infty \leq c_1$ and, when the left-hand
    sides are finite, $\vert m_i^a\vert \leq c_1$ and $\vert m_i^b\vert \leq c_1$.
  \end{assumption}

  The particular choice of norms in Assumption 1 is unimportant
  when $d$ is fixed but will matter in later sections when $d\to \infty$. To get
  some intuition for the first part of the assumption, consider for example the
  interval-censored regression in Example \ref{ex:int_cens_reg}. As noted
  following \eqref{eq:pmf_intcens}, when both $t_j^i$ and $t_{j + 1}^i$ are
  finite, $\bm{m}_i = \bm{0}$ and $\bm{\theta}^\tsp \bm{z}_i^a= \sigma^{-1}
  t^i_{j} - \bm{x}_i^\tsp \bm{\beta}/\sigma < \sigma^{-1} t^i_{j + 1} -
  \bm{x}_i^\tsp \bm{\beta} / \sigma$. Using this and that $\sigma_* > 0$, it is
  straightforward to show Assumption 1 holds (see Proof of
  Corolloray 1, Supporting Information, for an example). More generally,
  Assumption 1 ensures among other things that the support does
  not depend on $\bm{\theta}$ near $\bm{\theta}_*$.

  To state the first result, let $\emin(\cdot)$ denote the smallest eigenvalue
  of its matrix argument.

  \begin{theorem} \label{thm:main_small}
    If (a) $\mathcal{Y}$ is finite, (b) $r$ is strictly log-concave, strictly
    positive, and continuously differentiable on $\R{}$; (c) ${\bm{\theta}}_*$
    is an interior point of $\Theta$; (d) Assumption 1 holds;
    and (e)
    \begin{equation}\label{eq:cond_eig_finf}
          \liminf_{n\to \infty} \emin\left\{\sum_{i = 1}^n \E(\bm{Z}_i^\tsp \bm{Z}_i)
          / n\right\} > 0;
    \end{equation}
    then as $n\to \infty$ with $d$ fixed, $\bm{\mathcal{I}}_n(\bm{\theta}_*; \bm{X})^{1/2}(\widehat{{\bm{\theta}}}_n
      - {\bm{\theta}}_*) \rightsquigarrow \rN(\bm{0}, \bm{I}_d)$,
    where $\bm{\mathcal{I}}_n(\bm{\theta}_*; \bm{X}) = \cov\{\nabla
    \ell_n({\bm{\theta}}_*; \bm{Y}, \bm{X})\}$ is the Fisher information.
  \end{theorem}
  The proof of Theorem \ref{thm:main_small} uses a result by
  \citet{Hjort.Pollard2011} on minimizers of convex processes. The expectation
  and covariance in the theorem statement are with respect to the distribution
  of $\bm{Y} \mid \bm{X}$ under the true ${\bm{\theta}}_*$. In
  the proof it argued that $\bm{\mathcal{I}}_n(\bm{\theta}; \bm{X})$ has
  eigenvalues bounded below by $\epsilon n$ for some $\epsilon > 0$. With this,
  the theorem implies $\Vert\widehat{\bm{\theta}}_n - \bm{\theta}_* \Vert =
  O_{\pr}(1 / \sqrt{n})$. If $r$ is assumed to be twice
  continuously differentiable, then the conclusion of the theorem continues to
  hold if $\bm{\mathcal{I}}_n(\bm{\theta}; \bm{X})$ is replaced by the observed
  information $-\nabla^2 \ell_n(\bm{\widehat{\theta}}_n; \bm{Y}, \bm{X})$
  (Theorem B.5, Supporting Information). In the case of interval-censored
  linear regression, \eqref{eq:cond_eig_finf} reduces to a familiar condition on
  the design matrix $\bm{X} = [\bm{x}_1, \dots, \bm{x}_n]^\tsp \in \R{n \times
  p}$.

  \begin{corollary} \label{corol:int_cens_reg}
    Suppose $Y_1, \dots Y_n$ satisfy the interval-censored regression model in
    Example \ref{ex:int_cens_reg} with known $\sigma = 1$, $\mathcal{Y}$ is
    finite, the density of $W_i$ satisfies requirement (b) of Theorem
    \ref{thm:main_small}, and $\Vert \bm{x}_i\Vert \leq c_1$; then the
    conclusion of Theorem \ref{thm:main_small} holds if $\liminf_{n \to
    \infty}\emin\{\bm{X}^\tsp \bm{X} / n\} > 0$ as $n\to \infty$ with $d = p$
    fixed.
  \end{corollary}

  When $k\geq 2$ in the setting of Corollary \ref{corol:int_cens_reg}, we expect
  the conclusion can be shown to hold also when $\sigma$ is unknown.
  Intuitively, when the support of the response variables has cardinality
  greater than two, the variance need not be a function of the mean, and it may
  then be possible to estimate an additional parameter. By contrast, it is
  well-known $\sigma$ is unidentifiable in general in logistic and probit
  regression, which are special cases.

  \subsection{Diverging number of parameters} \label{sec:hd}
  
  Our second main result gives convergence rates for maximum $L_1$-regularized
  likelihood estimators when $d$ tends to infinity with $n$ and
  ${\bm{\theta}}_*$ is sparse. Since $d$ varies ${\bm{\theta}}_*$ generally
  depends on $d$, but we suppress this in notation. We consider the
  penalized average negative log-likelihood defined for $\lambda_n \geq 0$ by
  $G^\lambda_n({\bm{\theta}}; \bm{Y}, \bm{X}) =  G_n({\bm{\theta}}; \bm{Y}, \bm{X}) + \lambda_n \Vert
    {\bm{\theta}}\Vert_1$,
  and
  $
      \widehat{{\bm{\theta}}}_n^\lambda \in \argmin_{{\bm{\theta}} \in \Theta}
      G_n^\lambda({\bm{\theta}}; \bm{Y}, \bm{X}).
  $
  For any ${\bm{\theta}} \in \R{d}$ and $S \subseteq \{1, \dots, d\}$, define
  ${\bm{\theta}}_S \in \R{d}$ to equal ${\bm{\theta}}$ with the $j$th element
  set to zero if $j \notin S$, $j \in\{ 1, \dots, d\}$:
  \[
    ({\bm{\theta}}_{S})_j = \begin{cases} \theta_j & j \in S \\ 0 & j\notin S \end{cases}.
  \]
   We say ${\bm{\theta}}$ is $s$-sparse if ${\bm{\theta}} = {\bm{\theta}}_S$ for some $S \subseteq \{1,
   \dots, d\}$ with cardinality $s = \vert S\vert$.

  To state results, define the cone $
    \mathbb{C}(S) = \{{\bm{\theta}} \in \R{d}: \Vert {\bm{\theta}}_{S^c}\Vert_1 \leq 3 \Vert
    {\bm{\theta}}_S\Vert_1\}$,
  where $S^c = \{1, \dots, d\} \setminus S$ and, hence, ${\bm{\theta}}_{S^c} =
  {\bm{\theta}} - {\bm{\theta}}_S$. Intuitively, $\mathbb{C}(S)$ is a set of
  nearly-sparse ${\bm{\theta}}$ in the sense that the elements $\theta_j, j
  \notin S$, are not too large compared with the $\theta_j, j \in S$. Define
  also for any $\kappa > 0$, $n$, and $d$ the set
  \[
    \mathcal{C}_{\kappa, n, d} =  \left\{(\bm{Y}, \bm{X}) : \inf_{{\bm{\theta}} \in
    \mathbb{C}(S): \Vert {\bm{\theta}}\Vert = 1} \left\{{\bm{\theta}}^\tsp \left(\frac{1}{n}\sum_{i = 1}^n \bm{Z}_i^\tsp \bm{Z}_i
    \right) {\bm{\theta}} \right\} \geq \kappa \right\}.
  \]

  We are ready to state the next result.

\begin{theorem}\label{thm:hd}
  If (a) $\Theta$ is open, (b) $r$ is strictly log-concave, strictly positive,
  and continuously differentiable on $\R{}$; (c) ${\bm{\theta}}_*$ is $s$-sparse
  and $\Vert {\bm{\theta}}_*\Vert_\infty \leq c_1$; (d) Assumption
  1 holds; and (e) $\lambda_n = c_2 \log(d) / n  \to 0$; then
  there are $c_3, c_4, c_5$ such that, for large enough $n$ and $d$, with
  probability at least $\pr(\mathcal{C}_{\kappa, n, d})- d^{-c_3}$,
  \[
    \Vert \widehat{{\bm{\theta}}}^\lambda_n - {\bm{\theta}}_*\Vert^2 \leq c_4\frac{\log
    (d)}{n};\quad \Vert \widehat{{\bm{\theta}}}^\lambda_n - {\bm{\theta}}_*\Vert_1 \leq c_5
    \sqrt{\frac{ \log (d)}{n}}.
  \]
\end{theorem}
Assumptions (a) and (b) ensure the gradient and Hessian of $G_n$ exist. In some
settings of interest, for example interval-censored regressions with known
error variance, the matrices $\bm{Z}_1, \dots, \bm{Z}_n$ do not depend on
$\bm{Y}$. Then the event $\mathcal{C}_{\kappa, n, d}$ either contains all
outcomes or none and is hence better thought of as a restricted eigenvalue
condition on the deterministic $\sum_{i = 1}^n \bm{Z}_i^\tsp \bm{Z}_i / n$.
Specifically, if the inequality in the definition of $\mathcal{C}_{\kappa, n,
d}$ holds for some $\kappa > 0$ and all $n$ and $d$, and the other conditions of
the theorem hold, then the conclusion of the theorem holds with probability at
least $1 - d^{-c_3}$. Moreover, in interval-censored regression with known
error variance, a restricted eigenvalue condition on $\sum_{i = 1}^n
\bm{Z}_i^\tsp \bm{Z}_i / n$ is equivalent to one on $\bm{X}^\tsp \bm{X} / n$
since, in those cases, $\bm{Z}_i = -[\bm{x}_i, \bm{x}_i]^\tsp$, with one of the
rows replaced by zeros if $a_i(y_i, \bm{x}_i, {\bm{\theta}}) = -\infty$ or
$b_i(y_i, \bm{x}_i, {\bm{\theta}}) = \infty$.

It is common in the literature for the bounds on norms of
$\widehat{{\bm{\theta}}}^\lambda_n - {\bm{\theta}}_*$ to depend linearly on $s$
\citep[e.g.,][Corollary 2]{Negahban.etal2012}. Here, $s$ is fixed and absorbed
in the constants $c_4$ and $c_5$. This is because our proofs require $\bm{Z}_i
{\bm{\theta}}_*$ to be contained in a compact subset of $\R{2}$. We expect the
linear dependence on $s$ can be recovered in many special cases, though it may
require substantial work; see for example \citet{Negahban.etal2009} for the
special case of logistic regression.

\section{Computing} \label{sec:compute}
\subsection{Inexact proximal Newton}
We propose using an inexact proximal Newton algorithm for computing
$\widehat{{\bm{\theta}}}^\lambda_n$ in practice. That is, a proximal Newton
algorithm where the sub-problems are solved inexactly. Similar algorithms have
proven useful in, for example, the fitting of penalized generalized linear
models \citep{Lee.etal2006, Friedman.etal2010, Yuan.etal2012, Byrd.etal2016}.
The R package {\tt fsnet} (\citeyear{Thefsnetpackage2022})
implements the algorithm, and an accelerated proximal gradient descent algorithm similar to the
Fast Iterative Shrinkage-Thresholding (FISTA) algorithm
\citep{Beck.Teboulle2009}. We focus on the proximal Newton algorithm here
because we found it tends to perform well. It is often useful in practice
to include a ridge penalty and hence we solve the convex elastic-net
optimization problem
\begin{equation} \label{eq:elastic_net}
  \min_{{\bm{\theta}} \in \Theta} \left\{G_n({\bm{\theta}}; \bm{Y}, \bm{X}) + \lambda_{1}\Vert
  {\bm{\theta}}\Vert_1 + \frac{\lambda_2}{2} \Vert {\bm{\theta}}\Vert^2\right\},
\end{equation}
where $\lambda_1 \geq 0$ and $\lambda_2 \geq 0$ are user-specified penalty
parameters. The setting in Section \ref{sec:hd} is a special
case with $\lambda_2 = 0$. If $\lambda_2 > 0$ the objective function is
strongly convex and has a unique global minimizer. To simplify notation, let us
suppress dependence on the data $(\bm{Y}, \bm{X})$ for the remainder of the section
and re-define $G_n^\lambda$ to include the ridge penalty. That is,
$G_n^\lambda({\bm{\theta}})$ is the objective function in \eqref{eq:elastic_net}.

Proximal Newton solves \eqref{eq:elastic_net} by iteratively updating and
minimizing an $L_1$-penalized quadratic approximation of $G_n^\lambda$.
To be more specific, let $Q(\cdot; {\bm{\theta}}^k)$ denote a quadratic approximation of the
differentiable part of $G_n^\lambda$ at the $k$th iterate ${\bm{\theta}}^k$, given by
\[
  Q({\bm{\theta}}; {\bm{\theta}}^k) = \{\nabla G_n({\bm{\theta}}^k) + \lambda_2
  {\bm{\theta}}^k\}^\tsp {\bm{\theta}} + \frac{1}{2}({\bm{\theta}} - {\bm{\theta}}^k)^\tsp\{\nabla^2
  G_n({\bm{\theta}}^k) + \lambda_2 \bm{I}_d\}({\bm{\theta}} - {\bm{\theta}}^k).
\]
Then the $(k + 1)$th iterate in the proximal Newton algorithm is
\begin{equation} \label{eq:prox_update}
  {\bm{\theta}}^{k + 1} \approx \argmin_{{\bm{\theta}} \in \Theta} \left\{ Q({\bm{\theta}}; {\bm{\theta}}^k) +
  \lambda_1\Vert {\bm{\theta}}\Vert_1\right\},
\end{equation}
where $\approx$ indicates it is not necessary to solve the optimization problem
exactly (see Section \ref{sec:converge}). The update \eqref{eq:prox_update} does
not in general admit a closed form solution but can be solved efficiently to
desired tolerance using coordinate descent.

\subsection{Coordinate descent}

To discuss the coordinate descent algorithm for \eqref{eq:prox_update}, we
assume $\Theta = \R{d}$ for simplicity. Settings where some parameters need
to be positive (e.g., to ensure monotonic splines in an interval-censored
flexible parametric model) or not penalized (e.g., the error scale parameter in
an interval-censored regression), could be treated by minor modifications and
are supported in our software.

The $(l + 1)$th iterate for the $j$th component in a coordinate descent
algorithm for \eqref{eq:prox_update} is
\begin{equation}\label{eq:cd}
\theta^{k, l + 1}_j = \argmin_{\theta_j \in \R{}}\{Q([\theta_1^{k, l + 1},
  \dots, \theta_j, \theta_{j + 1}^{k, l}, \dots, \theta_d^{k, l}]^\tsp;
  \bm{\theta}^k) + \lambda_1 \vert \theta_j\vert\}.
\end{equation}
This is a univariate $L_1$-penalized quadratic optimization problem which can be
solved in closed form using the soft-thresholding operator. To be more specific,
define $\bm{g}:\R{2} \to \R{2}$ and $\bm{H}:\R{2} \to \R{2 \times 2}$ as,
respectively, the gradient and Hessian of the map $(t_1, t_2) \mapsto \log
\{R(t_2) - R(t_1)\}$, $-\infty < t_1 < t_2 <\infty$. Extend also $\bm{g}$ and
$\bm{H}$ to include points where $t_1 = -\infty$ by setting the first element of
$\bm{g}$ and first row and column of $\bm{H}$ to zero at such points. Similarly,
extend to points with $t_2 = \infty$ by setting the second element of $\bm{g}$
and second row and column of $\bm{H}$ to zero at such points.  Then
\[
  \nabla G_n({\bm{\theta}}; \bm{Y}, \bm{X}) = -\frac{1}{n}\sum_{i = 1}^n \bm{Z}_i^\tsp \bm{g}(\bm{Z}_i {\bm{\theta}} + \bm{m}_i);~~
  \nabla^2 G_n({\bm{\theta}}; \bm{Y}, \bm{X}) = -\frac{1}{n}\sum_{i = 1}^n \bm{Z}_i^\tsp \bm{H}(\bm{Z}_i
  {\bm{\theta}}+ \bm{m}_i) \bm{Z}_i.
\]
Let $\bm{\eta}_i^k = \bm{Z}_i {\bm{\theta}}^k + \bm{m}_i$, $\bm{H}_i^k = \bm{H}(\bm{\eta}_i^k)$, $\bm{g}_i^k = \bm{g}(\bm{\eta}_i^k)$,
and
\[
\bm{\eta}_{i(j)}^{k, l} = \sum_{m < j}\bm{Z}_i^m\theta_m^{k, l + 1} + \sum_{m > j} \bm{Z}_i^m\theta_m^{k, l} + \bm{m}_i,
\]
where $\bm{Z}_i^m$ is the $m$th column of $\bm{Z}_i$. Up to terms not depending
on $\theta_j$, the objective function in \eqref{eq:cd} is
\[
  \frac{\theta_j}{n}\sum_{i = 1}^n \{-(\bm{Z}_i^j)^\tsp \bm{g}_i^k + \lambda_2
  \theta_j^k + (\bm{\eta}_i^k - \bm{\eta}_{i(j)}^{k, l})^\tsp \bm{H}_i^k
  \bm{Z}_i^j\} + \frac{\theta_j^2}{2n} \sum_{i = 1}^n \{-(\bm{Z}_i^j)^\tsp
  \bm{H}_i^k \bm{Z}_i^j + \lambda_2\} + \lambda_1\vert \theta_j\vert.
\]
Using this, a routine calculation shows the minimizer in \eqref{eq:cd} is
\begin{equation} \label{eq:soft_solution}
  \theta_j^{k, l + 1} = \frac{\operatorname{soft}\left[-\frac{1}{n}\sum_{i =
  1}^n \{-(\bm{Z}_i^j)^\tsp \bm{g}_i^k + \lambda_2 \theta_j^k +   (\bm{\eta}_i^k - \bm{\eta}_{i(j)}^{k, l})^\tsp \bm{H}_i^k \bm{Z}_i^j\} , \lambda_1\right]}{\frac{1}{n}\sum_{i = 1}^n
  \{-(\bm{Z}_i^j)^\tsp \bm{H}_i^k \bm{Z}_i^j + \lambda_2\}},
\end{equation}
where $\operatorname{soft}(x, \lambda) = \mathrm{sign}(x)\max\{\vert x\vert -
\lambda, 0\}$. Notably, $\bm{\eta}_{i(j)}^{k, l}$ is the only term that needs
updating in the coordinate descent, making each step fast to compute.

The resulting algorithm is stated in Algorithm \ref{alg:prox_new}.

\begin{algorithm}
\caption{Proximal Newton with coordinate descent}
\label{alg:prox_new}
\begin{enumerate}
\item Input $\lambda_1 \geq 0, \lambda_2 \geq 0$, ${\bm{\theta}}^1 \in \R{d}$
\item For $k = 1 , 2, \dots$ until convergence:
\begin{enumerate}
\item Let ${\bm{\theta}}^{k, 1} = {\bm{\theta}}^k$ and for $l = 1, 2, \dots$ until convergence, update $\theta^{k, l + 1}_j$ iteratively for $j = 1, \dots, d$ according to \eqref{eq:soft_solution}.
\item Let ${\bm{\theta}}^{k, l}$ be the vector of final iterates
from (a) and set ${\bm{\theta}}^{k + 1} = (1 - s){\bm{\theta}}^k + s {\bm{\theta}}^{k, l}$ with $s \in
[0, 1]$ selected by backtracking line-search.
\end{enumerate}
\item Return final iterate ${\bm{\theta}}^{k + 1}$ from 2.
\end{enumerate}
\end{algorithm}

\subsection{Convergence} \label{sec:converge}

Convergence of Algorithm \ref{alg:prox_new} can be guaranteed by selecting
appropriate termination criteria for the inner coordinate descent algorithm
(step 2 (a)) and the backtracking line-search (step 2 (b)). It will be
convenient to characterize solutions to \eqref{eq:elastic_net} using the
function $\bm{J}: \Theta \to \R{d}$ defined for $c_1 > 0$ by
$
  \bm{J}({\bm{\theta}}; c_1) = \nabla G_n({\bm{\theta}}) + \lambda_2 {\bm{\theta}} -
  \mathcal{P}_{\lambda_1}\{\nabla G_n({\bm{\theta}}) + \lambda_2 {\bm{\theta}} -
  {\bm{\theta}} / c_1\},
$
where $\mathcal{P}_{\lambda_1}$ is the elementwise projection onto $[-\lambda_1,
\lambda_1]$. Routine calculations show, for any $c_1 > 0$, $\bm{J}({\bm{\theta}}; c_1) = 0$
if and only if $0$ is a sub-gradient of $G^\lambda_n$ at ${\bm{\theta}}$
\citep{Milzarek.Ulbrich2014,Byrd.etal2016}; that is, $\bm{J}({\bm{\theta}}; c_1) = 0$ if and
only if ${\bm{\theta}}$ is a solution to \eqref{eq:elastic_net}. Similarly, ${\bm{\theta}}$ is
a solution to \eqref{eq:prox_update} if and only if $\bm{J}_Q({\bm{\theta}}; c_1, {\bm{\theta}}^k)
= 0$, where
\begin{align*}
  \bm{J}_Q({\bm{\theta}}; c_1, {\bm{\theta}}^k) &= \nabla G_n({\bm{\theta}}^k) + \lambda_2 {\bm{\theta}}^k +
  \{\nabla^2 G_n({\bm{\theta}}^k) + \lambda_2 \bm{I}_d\}({\bm{\theta}} - {\bm{\theta}}^k) \\
  &\quad  - \mathcal{P}_{\lambda_1}[\nabla G_n({\bm{\theta}}^k) + \lambda_2 {\bm{\theta}}^k +
  \{\nabla^2 G_n({\bm{\theta}}^k) + \lambda_2 \bm{I}_d\}({\bm{\theta}} - {\bm{\theta}}^k) - {\bm{\theta}} / c_1].
\end{align*}
Following \citet{Byrd.etal2016}, the coordinate descent algorithm
for \eqref{eq:cd} may be terminated when the $l$th coordinate descent iterate
${\bm{\theta}}^{k, l} = [\theta_1^{k, 1}, \dots, \theta_{d}^{k, l}]^\tsp$ satisfies,
for a user-specified $c_2 \in [0, 1)$,
\begin{equation}\label{eq:terminate_cd}
  \Vert \bm{J}_Q({\bm{\theta}}^{k, l}; {\bm{\theta}}^k, c_1)\Vert \leq c_2 \Vert \bm{J}_Q({\bm{\theta}}^{k};
  {\bm{\theta}}^k, c_1)\Vert.
\end{equation}

To specify a termination criterion for the line-search in step 2 (b), define a
first-order approximation of $G_n^\lambda$ at ${\bm{\theta}}^k$ by
\[
L_n^\lambda({\bm{\theta}}; {\bm{\theta}}^k) = G_n({\bm{\theta}}^k) + 0.5 \lambda_2\Vert
{\bm{\theta}}^k\Vert^2 + \{\nabla G_n({\bm{\theta}}^k) + \lambda_2 {\bm{\theta}}^k\}^\tsp({\bm{\theta}} -
{\bm{\theta}}_k) + \lambda_1 \Vert {\bm{\theta}}\Vert_1.
\]
Given ${\bm{\theta}}^{k, l}$ satisfying \eqref{eq:terminate_cd}, backtracking
line-search starts with step-size $s = 1$ and decreases until, for a
user-specified $c_3 \in (0, 1/2)$,
\begin{equation}\label{eq:backtrack}
  G_n^{\lambda}({\bm{\theta}}^k) - G_n^{\lambda}\{(1 - s){\bm{\theta}}^{k} + s {\bm{\theta}}^{k, l}\}
  \geq c_3 [L_n^\lambda({\bm{\theta}}^k; {\bm{\theta}}^k) -
L_n^\lambda\{(1 - s){\bm{\theta}}^{k} + s {\bm{\theta}}^{k, l}; {\bm{\theta}}^k\}].
\end{equation}

We are ready to state a convergence result for Algorithm \ref{alg:prox_new}.

\begin{theorem} \label{thm:comp}
  If in Algorithm \ref{alg:prox_new} convergence in step 2 (a) is determined
  using \eqref{eq:terminate_cd}, the backtracking linesearch in step 2 (b)
  satisifes \eqref{eq:backtrack}, $r$ is continuously differentiable, and either
  \begin{itemize}
  \item[(a)] $r$ is strictly log-concave and strictly positive, $\sum_{i = 1}^n
  \bm{Z}_i^\tsp \bm{Z}_i$ is positive definite, and $\inf_{{\bm{\theta}} \in \Theta}
  G^\lambda_n({\bm{\theta}}) = G^\lambda_n(\widehat{{\bm{\theta}}})$ for some $\widehat{{\bm{\theta}}} \in \Theta$; or
  \item[(b)] $\lambda_2 > 0$;
\end{itemize}
   then the sequence $\{{\bm{\theta}}^k\}$ of iterates
   satisfies $\lim_{k \to \infty} \bm{J}({\bm{\theta}}^k; c_1) = 0$.
\end{theorem}

Conditions (a) and (b) are used to show, among other things, the iterates
$\{{\bm{\theta}}_k\}$ stay in a compact set. If this can be guaranteed by other
means, some conditions can be weakened. For example, it is typically possible to
relax the first two requirements in (a) if the gradient is Lipschitz-continuous
and the Hessian in the quadratic approximation $Q({\bm{\theta}};
{\bm{\theta}}^k)$ is regularized to have eigenvalue bounded away from zero.
Notably, we have had no convergence issues in simulations even when $\lambda_2 =
0$ and $\sum_{i = 1}^n \bm{Z}_i^\tsp \bm{Z}_i$ is indefinite because $d > n$, as
long as $\lambda_1 > 0$.

\section{Numerical experiments}

We illustrate the proposed methods in two interval-censored regression models
(see Example \ref{ex:int_cens_reg}). In the first, $W$ has the extreme-value
distribution with cumulative distribution function $R(w) = 1 -
\exp\{-\exp(w)\}$, and the number of predictors $p = 3$ is smaller than the
number of observations $n = 100$. When $W$ has the extreme-value distribution in
\eqref{eq:int_cens_reg} and $\sigma = 1$, $\exp(Y^*)$ has the exponential
distribution with mean $\exp(\bm{x}^\tsp {\bm{\theta}})$. This model is a
special case of that in Example \ref{ex:flex_par_surv}. It is also a special case
of a gamma generalized linear model with logarithm link function, which we
therefore include in comparisons. The observed response indicates whether
$\exp(Y^*)$ is in $[0, d), [d, 2d), \dots, [kd, 5)$, or $[5, \infty)$, where $d$
(the interval size) varies in the simulations and $k$ is the largest integer
such that $kd < 5$. Thus, a larger $d$ corresponds to more severe censoring.
Because $Y^*$ is not observed, when fitting the generalized linear model we take
the upper endpoints of the observed intervals, or $5 + d$ if the interval is
$[5, \infty)$, as responses.

In the second setting $W$ is normally distributed and $p = 200 > 100 = n$. The
observed intervals are for $Y^*$ are $(-\infty, 5), [5, -kd), \dots, [-d, 0),
[0, d), \dots, [kd, 5), [5, \infty)$. We compare the estimates from Algorithm
\ref{alg:prox_new}  with $\lambda_2 = 0$ to those from lasso regression using
{\tt glmnet} \citep{Friedman.etal2010}. For both methods, the regularization
parameter $\lambda_1$ is selected by 5-fold cross-validation. For our method, we
select the $\lambda_1$ which minimizes the average out-of-sample
misclassification rate. Here, the misclassification rate for one fold is the
proportion of observations $(Y_i, \bm{x}_i)$ in that fold for which the
predicted mean $\bm{x}_i^\tsp \widehat{{\bm{\theta}}}_n^\lambda$ of the $i$th
unobservable response $Y_i^*$ is outside the observed interval.

The predictors are generated as centered and scaled realizations from a
multivariate normal distribution with mean zero and a covariance matrix with
$(i, j)$th element $0.5^{\vert i - j\vert}$. When $p = 3$ we include an
intercept so there are two jointly normal predictors in addition to the
intercept. The true coefficient vector is ${\bm{\theta}}_* = [1, 1/2,
-1/2]^\tsp$ when $p = 3$ and ${\bm{\theta}}_* = [1,  1/2, -1/2, 0, \dots,
0]^\tsp$ when $p = 200$.

\begin{figure}[h]
\centering
\includegraphics[width = \textwidth]{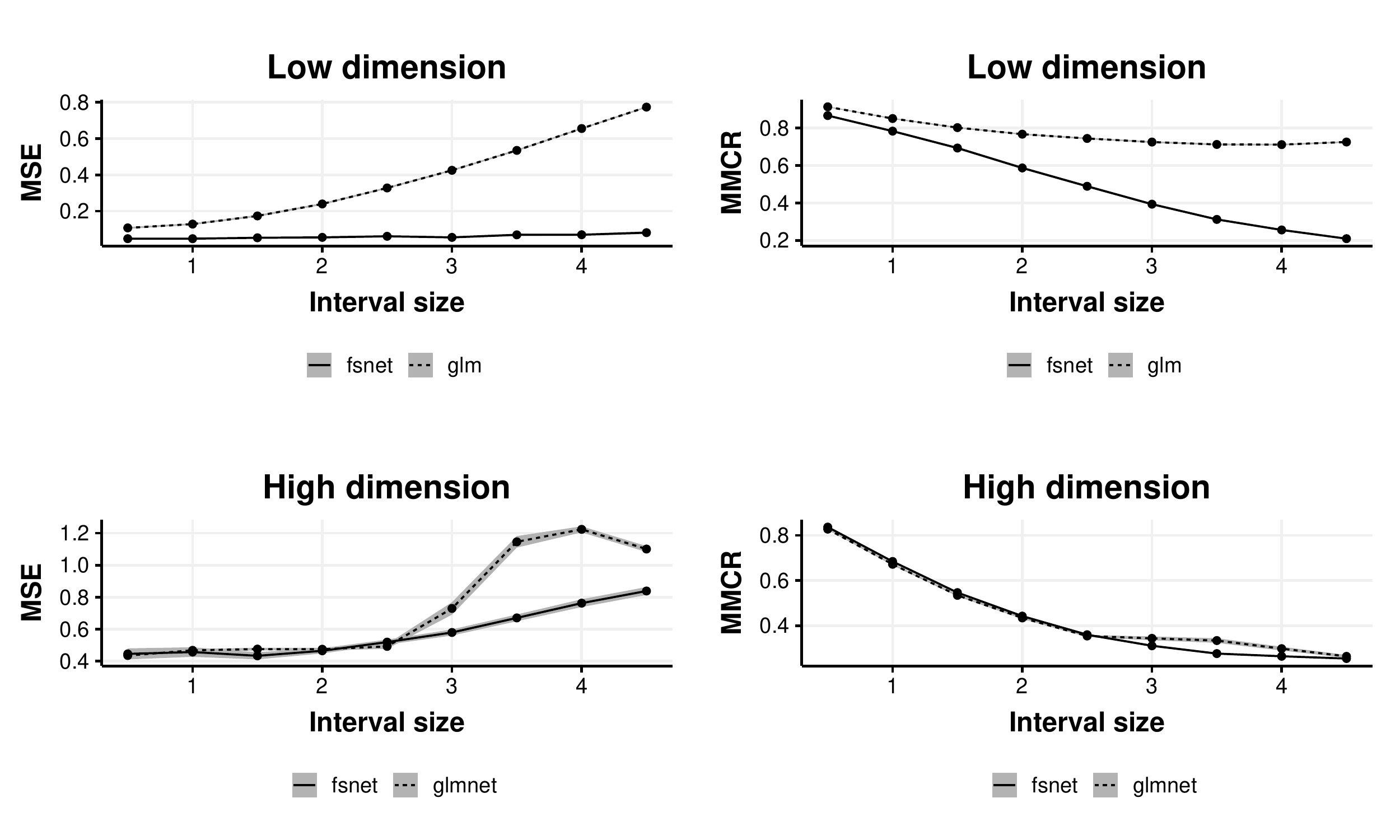}
\caption{Sum of squared estimation errors and mean misclassification rates for
the proposed method (fsnet), generalized linear models (glm), and lasso
regression (glmnet). The shaded confidence bands are $\pm 1.96$
times the Monte Carlo standard errors.}
\label{fig:sims}
\end{figure}

Figure \ref{fig:sims} shows how sum of squared estimation errors for the
non-zero components of ${\bm{\theta}}_*$ and mean misclassification rates vary with the
interval size $d$. The sum of squared estimation errors is defined as $\sum_{j =
1}^3 \sum_{i = 1}^{m} (\widehat{{\bm{\theta}}}^i_{j} - {\bm{\theta}}_{*j})^2$, where $m = 500$ is
the number of replications in the simulations and $\widehat{{\bm{\theta}}}_j^i$ is an
estimate of the $j$th element of ${\bm{\theta}}_*$ in the $i$th replication.

The first row of Figure \ref{fig:sims} shows, as expected, using the correct
likelihood is beneficial, and the benefits are greater the more severe the
interval-censoring. We note the mean misclassification rate for the generalized
linear model decreases as the interval-censoring gets more severe, which is an
effect of it being easier to predict the correct interval when the intervals
are larger.

The second row in Figure \ref{fig:sims} indicates the proposed method can, when
intervals are small enough, perform similarly to that based on the incorrect
normal likelihood; that is, to lasso regression. Some intuition for this can be
gained by considering the bias-variance trade-off in estimating
${\bm{\theta}}_*$: bias is introduced by using the incorrect likelihood, but if
the intervals are small enough that bias is small in comparison to the variance.
Indeed, the large variance in high-dimensional settings is a key reason
regularization, which introduces bias but decreases variance, is often useful.
As the censoring becomes more severe, however, the bias is again substantial.

\section{Data examples} \label{sec:data_ex}

\subsection{Lipoprotein data} \label{sec:plasma}

Lipoprotein(a) [Lp(a)] is a risk factor for cardiovascular complications (see
for example \citet{Littmann.etal2019} or \citet{Littmann.etal2022}). Hence, it
is of interest to model the distribution of Lp(a) in different populations and
to investigate the effects of covariates. One challenge is that Lp(a) has a
lower limit of detection of 10 nanomoles per liter (nmol / L), leading to
censoring from below. Additionally, in practice it Lp(a) is often categorized,
into classes, such as those defined by deciles. We consider a
regression model for Lp(a) in nmol / L in intervals $[0, 10], (10, 20], \dots,
(110, 120], (120, \infty)$. The data are a subset of those used by
\citet{Littmann.etal2019}, except they use different classes, $[0, 10], (10, 30],
(30, 120], (120, \infty)$. There are $n = 1837$ observations and four covariates:
sex, age, smoking status (never-smoker, ex-smoker, or smoker), and hemoglobin
A1c (HbA1c) measurements categorized into three levels (low, average, high)
corresponding to good, average, and poor metabolic control.

We first fit a model for Lp(a) without predictors. One possibility is to fit the
cumulative probability model in Example \ref{ex:categorical}. As argued there,
this is equivalent to fitting a general categorical model with $12$ parameters,
the number of categories minus one. For concreteness, take $R$ to be the
standard normal cumulative distribution function and denote the maximum
likelihood estimate by $\widehat{\bm{\theta}}^c$. This estimate ensures
$R(\widehat{\theta}^c_j) = n^{-1}\sum_{i = 1}^n Y_i \I(Y_i \leq 10 j), j \in\{
1, \dots, 12\}$, where $\I(\cdot)$ is an indicator function. That is, the
estimated category probabilities equal the sample proportions.

\begin{figure}
  \centering
  \includegraphics[width = 0.8\textwidth]{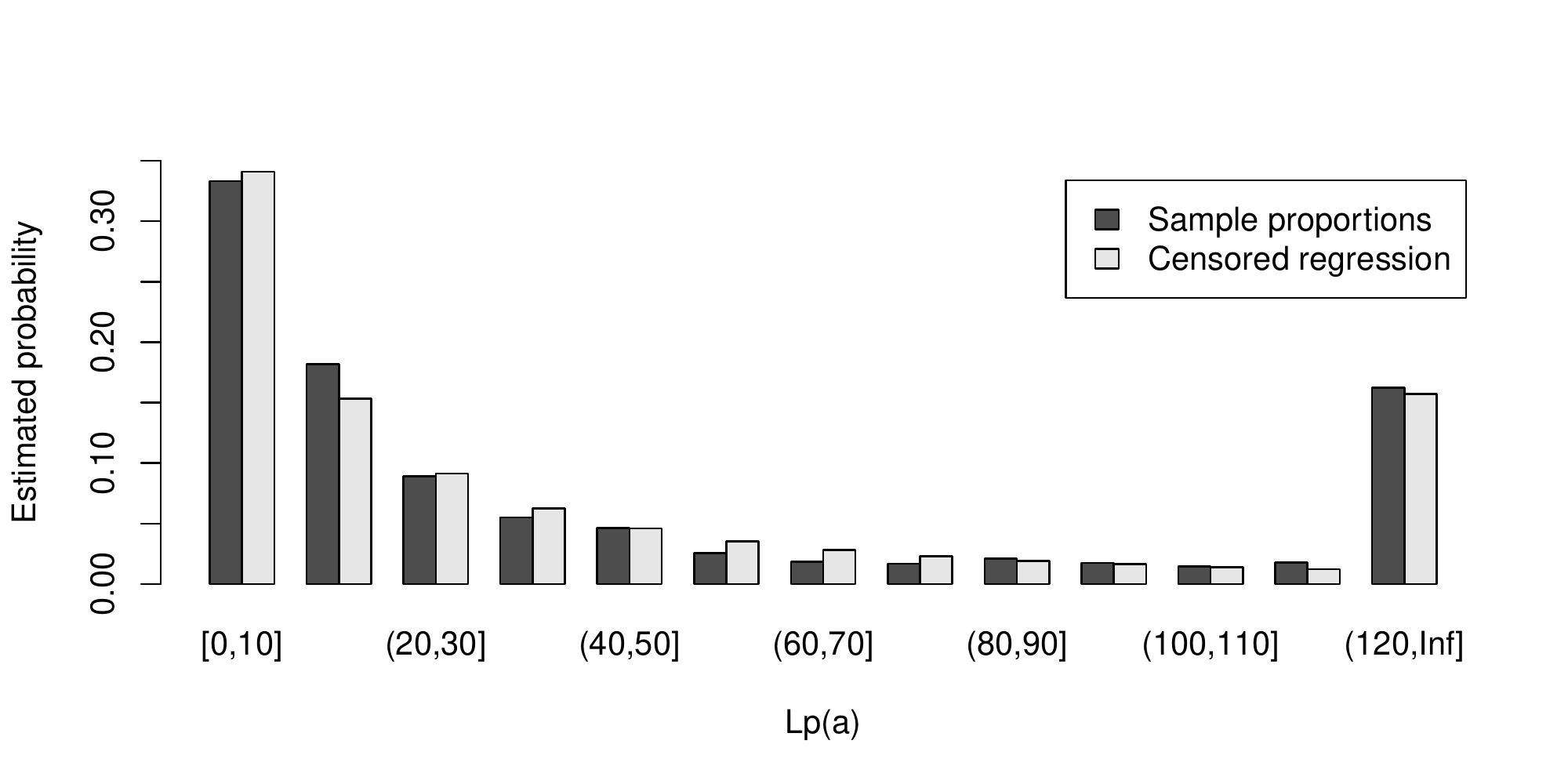}
  \caption{Estimated probability mass function for censored Lp(a)}
  \label{fig:pdf_plasma}
\end{figure}

Another possibility is to assume the Lp(a) measurements come from a censored
regression model such as in Example \ref{ex:int_cens_reg} with an intercept
only. This can be particularly useful when interest is in inference on the
unobservable continuous Lp(a). Because the model for the unobservable continuous
Lp(a) is the same regardless of the censoring, this model facilitates pooling
data from studies with different censoring. Since Lp(a) must be positive, we
consider the model which assumes the continuous Lp(a), $Y_i^*$, satisfies
\[
    \log(Y_i^*) = \beta + \sigma W_i,
\]
where $W_i \sim \rN(0, 1)$. Our response $Y_i$ is the interval containing
$Y_i^*$. Fitting this model we get the maximum likelihood estimates $\hat{\beta}
= 3.02$ and $\hat{\sigma} = 1.76$. These can be interpreted as usual in the
latent regression, or one can focus on the estimated mass function for $Y_i$
given in Figure \ref{fig:pdf_plasma}. Notably, the estimated probabilities are
reasonably close to the sample proportions, or equivalently, the estimates from
the cumulative probability model.

The maximized likelihood for the latent regression model will always be lower
than that of the cumulative probability model since the latter is equivalent to
a general categorical model (Example \ref{ex:categorical}). However, the former
has fewer parameters and so may still be preferable. For example, computing the
BIC for both models shows the smaller is preferable with a BIC of 7432 compared
with 7480. We also considered letting $W_i$ have an extreme-value distribution,
but that gave a BIC of 7481.

To investigate the effect of covariates, we continue with a latent regression
model: $\log(Y_i^*) = \bm{x}_i^\tsp \bm{\beta} + \sigma  W_i$, $W_i \sim
\rN(0, 1)$. Following \citet{Littmann.etal2019}, we consider the effect of age
on Lp(a), and whether there are interactions between age and the other
covariates. Considering the interactions first, we compare two models using a
likelihood ratio test, a smaller one where
\[
    \bm{x}_i = [1, \mathtt{age}_i, \mathtt{male}_i,
    \mathtt{never\_smoker}_i, \mathtt{smoker}_i, \mathtt{average\_hba1c}_i,
    \mathtt{high\_hb1ac}_i]^\tsp,
\]
and a larger one where $\bm{x}_i$ also includes age interacted with all the
other predictors. The likelihood ratio test with 5 degrees of freedom gave a
$p$-value of 0.19, indicating the interactions are not important.

Coefficient estimates and standard errors based on the observed information for
the smaller model are in Table \ref{tab:cens_regr}. The reported $p$-values are
for Wald-type tests for whether a regression coefficient is zero and whether the
scale parameter $\sigma = 1$. Any $p$-value less than $10^{-4}$ is reported as $0$. In summary, there is evidence Lp(a) increases with
age and is associated with poor metabolic control.

\begin{table}[ht]

\centering
\begin{tabular}{r c c c c c c c c}
  & Scale & Int. & Age & Male & Nev. Smoker & Smoker & Med. HbA1c & High HbA1c \\
 \hline
Est. & 1.74 & 2.6 & 0.011 & -0.13 & -0.20 & -0.41 & 0.16 & 0.36 \\
S.E. & 0.046 & 0.21 & 0.0028 & 0.088 & 0.12 & 0.16 & 0.11 & 0.13 \\
$p$-value & $ 0$ & $0$ & $ 0$ & 0.13 & 0.082 & 0.010 & 0.14 & 0.0067\\
\hline
\end{tabular}
\caption{Regression for censored Lp(a)} \label{tab:cens_regr}
\end{table}

\subsection{Breast cancer data} \label{sec:cancer}

We use data from the Netherlands Cancer Institute on $n = 144$ lymph node
positive women \citep{vandeVijver.etal2002}. Following \citet[Examples 7.1 and
7.2]{Tutz.Schmid2016}, we model the time to development of distant metastases or
death, in three-month intervals up to 15 months. For each patient the
data include a follow-up time and an event indicator. The observable intervals
are $[0, 3), \dots, [12, 15)$, or one of those intervals with the upper
endpoint replaced by $\infty$ if the event (death or distant metastases) was not
observed.

The data also include five clinical predictor variables (diameter of tumor $>$ 2
cm or not, number of affected lymph nodes $\leq$ 3 or not, estrogen receptor
status positive or negative, tumor grade in three levels, and age) and gene
expression measurements for 70 genes. We first consider a model using the
clinical variables only, and then investigate whether the gene expression data
can be used to improve out-of-sample predictions.

Suppose, as in Example \ref{ex:flex_par_surv}, the
continuous, unobservable time-to-event $T_i$ has cumulative distribution
function
$
F(t_i; \bm{x}_i, {\bm{\beta}}, {\bm{\gamma}}) = 1 - \exp[-\exp\{\mathtt{sp}(\log t_i;
{\bm{\gamma}})-{\bm{\beta}}^\tsp \bm{x}_i\}],
$
where $\mathtt{sp}$ is a spline function. Specifically, we pick the I-splines
discussed by \citet{Ramsay1988} with no knots and three degrees of freedom,
implemented in the {\tt R} package {\tt splines2} \citep{Wang.Yan2021}. These
splines are monotone if the elements of ${\bm{\gamma}}$ are non-negative, which
we therefore enforce when fitting. Exponential and Weibull interval-censored
models are special cases corresponding to, respectively, $\mathtt{sp}(\log t;
\gamma) = \log t$ and $\mathtt{sp}(\log t; \gamma) = \gamma \log t$. The three
models are nested and upon fitting and comparing them using likelihood ratio
tests, we got the $p$-value 0.83 when testing the flexible I-splines against
Weibull, 0.88 for Weibull against exponential, and 0.53 for the flexible I-splines against
the exponential.

\begin{figure}[ht]
\centering
\includegraphics[width = \textwidth]{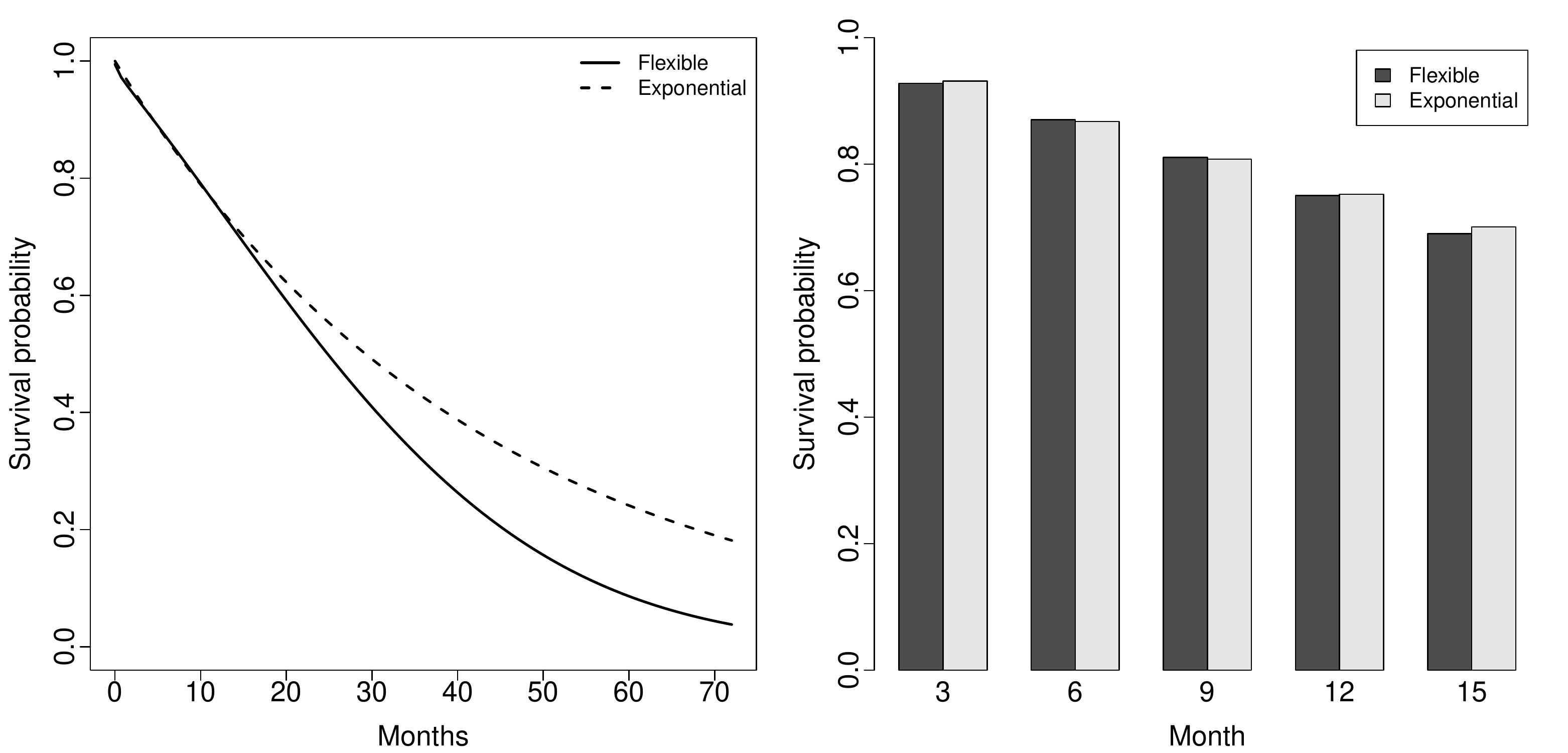}
\caption{Estimated survival probabilities for breast cancer data}
\label{fig:bc}
\end{figure}

Figure \ref{fig:bc} shows estimated survival probabilities for the flexible
I-splines and exponential models. In the figure, the clinical predictors are
held at their median values. The first plot shows a marked difference in
estimated survival probabilities in the right tail for the unobservable,
continuous survival times. However, for the observable data only the
probabilities at months $3, 6, \dots, 15$ matter. Indeed, any two survival
functions that agree at those points give the same distribution for the observed
data. The second plot in Figure \ref{fig:bc} shows the two models give similar
survival probabilities at the relevant points, consistent with the large
$p$-values obtained when comparing the different models. We focus on the
exponential model for the remainder of the section.

Table \ref{tab:exp_regr} shows results from fitting the exponential model. The
reported standard errors are square roots of diagonal entries of the inverse of
the observed Fisher information matrix. The $p$-values are Wald-type and are for
the null hypotheses that coefficients are zero. The number of affected lymph
nodes appears to be an important predictor, and there is some evidence the tumor
grade may be important.

\begin{table}[ht]

\centering
\begin{tabular}{r c c c c c c c}
  & Intercept & Diam. $>$ 2 & Nodes $\leq 3$ & E.R. Pos. & Grade.L & Grade.Q & Age \\
 \hline
Est. & 0.00072 & -0.30 & 0.77 & 0.58 & 0.55 & 0.26 & 0.051 \\
S.E. & 1.1 & 0.33 & 0.34 & 0.36 & 0.33 & 0.26 & 0.028\\
$p$-value & 1.0 & 0.35 & 0.022 & 0.11 & 0.098 & 0.33 & 0.068 \\
\hline
\end{tabular}
\caption{Regression for three-month time-to-event} \label{tab:exp_regr}
\end{table}

We next consider prediction using the gene expression measurements. Let $\bm{h}_i \in
\R{70}$ be a vector of gene expression measurements, standardized to have sample
mean zero and unit sample variance. We are interested in whether the $\bm{h}_i$ can
be used to improve the predictive performance of our method, and if so,
selecting genes useful for that purpose. To investigate we randomly split the
data into a test set of $144 / 3 = 48$ observations and a training set of $96$
observations. We consider the exponential model with predictor vector
$[\bm{x}_i^\tsp, \bm{h}_i^\tsp]^\tsp \in \R{77}$ and coefficient vector
${\bm{\theta}} = [{\bm{\beta}}^\tsp, \bm{\alpha}^\tsp]^\tsp \in \R{77}$, so $\bm{\alpha}
\in \R{70}$ is the coefficient vector for the gene expressions. Consider the
estimators
$
  \widehat{{\bm{\theta}}} = (\widehat{{\bm{\beta}}}, \widehat{\bm{\alpha}}) \in \argmin_{{\bm{\theta}} \in \R{77}}
  \{G_{n_t}({\bm{\theta}}) + \lambda \Vert \bm{\alpha}\Vert_1 \}
$
and
$
  \widetilde{{\bm{\beta}}} \in \argmin_{{\bm{\beta}} \in \R{7}} G_{n_t}([{\bm{\beta}}^\tsp, 0]^\tsp),
$
where $n_t = 96$ is the number of observations in the training set. The former
estimator penalizes the coefficients for the gene expression measurements while
the latter assumes those coefficients are zero. Thus, $\widetilde{{\bm{\beta}}}$ is the
maximum likelihood estimator in the exponential model without
gene expressions, using the training set only.

The penalty parameter $\lambda$ was selected from the set $\{\exp(-10), \dots,
\exp(0)\}$ by five-fold cross-validation on the training set. This gave $\lambda
= \exp(-5) \approx 0.0067$, which attained an average misclassification rate of
0.29 over the five folds. Predictions on the test-set with the selected
$\lambda$ gave an out-of-sample misclassification rate of 0.31. By comparison,
using the clinical predictors only, that is, the predictions $\exp(\bm{x}_i^\tsp
\widetilde{{\bm{\beta}}})$, gave a misclassification rate of 0.44. We conclude the
gene expression measurements can improve prediction, agreeing with the findings
of \citet{Tutz.Schmid2016}.

With $\lambda = \exp(-5)$, 32 of the 70 elements of $\widehat{\bm{\alpha}}$ were
zero. The Supporting Information contains a trace plot showing how the number of
non-zero coefficients and their sizes vary with $\lambda$.

\section{Conclusion}

The fact that observed data have finite support ought to be considered before
using models for continuous random variables, which in general leads to
misspecification bias. Roughly speaking, the smaller the cardinality of the
support and the variance of maximum likelihood estimators are, the more
pronounced the misspecification bias is. Even in settings where the bias is
small, however, the effects of using a misspecified likelihood can be difficult
to assess, leading to unreliable inference. With the methods proposed here
practitioners have access to fast and reliable likelihood-based inference, in
both low- ($n < d$) and high-dimensional ($d > n$) regression problems. There is
a wide range of possible applications, including but not limited to survival
analysis in discrete time, ordinal regression, and interval-censored linear
regression. Moreover, while the presented theory made repeated use of the
concavity of the log-likelihood, \eqref{eq:model} gives a valid model even if
$r$ is not log-concave. Thus, the modeling framework can be extended to many
settings not discussed in the present paper.

Possible directions for future research include the development of theory for
the interplay between the severity of censoring and the properties of maximum
likelihood estimators. For example, it may be informative to consider
asymptotics where the length of the censoring intervals is allowed to change
with the sample size and the number of parameters. Additionally, several special
cases of the models considered herein are also of significant interest in their
own right, and may hence merit further study. As noted in Section \ref{sec:asy},
more informative high-dimensional convergence bounds can likely be obtained for
special cases. It may also be worthwhile to explore settings with dependent
data. In the present setting, some types of dependent responses may be analyzed
by joining their supports. For example, two dependent binary responses can be
recoded as one response with four possible outcomes. The present setting could
also in principle be extended to include random effects in the linear
predictors, but the theory and implementation would require substantial work.

\section*{\centering Acknowledgements}
We thank Aaron Molstad for helpful discussions and Jonatan Risberg for
contributions to the software implementing the proposed methods. We are grateful
for comments from two reviewers and an Associate Editor which led to significant
improvements.

% \section*{\centering Data availability statement}
% The data that support the findings in Section 6.1 are not openly available as
% the authors do not have permission to share it.
% The data that support the findings in Section 6.2 are openly available in the R package {\tt penalized} (\citeyear{Thepenalizedpackage2022}). The data that support the findings in Web Appendix A.1 are available in the Supporting Information. The data that support the findings in Web Appendix A.2 are openly available in the R package {\tt glmnetcr} (\citeyear{Theglmnetcrpackage2022}).

\bibliographystyle{apalike}
\bibliography{finite_regr.bib}

\begin{thebibliography}{}

\bibitem[Agresti, 2019]{Agresti2019}
Agresti, A. (2019).
\newblock {\em An {{Introduction}} to {{Categorical Data Analysis}}}.
\newblock Wiley Series in Probability and Statistics. {John Wiley \& Sons},
  {Hoboken, NJ}, third edition edition.

\bibitem[Beck and Teboulle, 2009]{Beck.Teboulle2009}
Beck, A. and Teboulle, M. (2009).
\newblock A fast iterative shrinkage-thresholding algorithm for linear inverse
  problems.
\newblock {\em SIAM Journal on Imaging Sciences}, 2(1):183--202.

\bibitem[Burridge, 1981]{Burridge1981}
Burridge, J. (1981).
\newblock A note on maximum likelihood estimation for regression models using
  grouped data.
\newblock {\em Journal of the Royal Statistical Society. Series B
  (Methodological)}, 43(1):41--45.

\bibitem[Burridge, 1982]{Burridge1982}
Burridge, J. (1982).
\newblock Some unimodality properties of likelihoods derived from grouped data.
\newblock {\em Biometrika}, 69(1):145--151.

\bibitem[Byrd et~al., 2016]{Byrd.etal2016}
Byrd, R.~H., Nocedal, J., and Oztoprak, F. (2016).
\newblock An inexact successive quadratic approximation method for {{L-1}}
  regularized optimization.
\newblock {\em Mathematical Programming}, 157(2):375--396.

\bibitem[Couso et~al., 2017]{Couso.etal2017}
Couso, I., Dubois, D., and H{\"u}llermeier, E. (2017).
\newblock Maximum {{Likelihood Estimation}} and {{Coarse Data}}.
\newblock In Moral, S., Pivert, O., S{\'a}nchez, D., and Mar{\'i}n, N.,
  editors, {\em Scalable {{Uncertainty Management}}}, Lecture {{Notes}} in
  {{Computer Science}}, pages 3--16, {Cham}. {Springer International
  Publishing}.

\bibitem[Finkelstein, 1986]{Finkelstein1986}
Finkelstein, D.~M. (1986).
\newblock A proportional hazards model for interval-censored failure time data.
\newblock {\em Biometrics}, 42(4):845.

\bibitem[Friedman et~al., 2010]{Friedman.etal2010}
Friedman, J.~H., Hastie, T., and Tibshirani, R. (2010).
\newblock Regularization paths for generalized linear models via coordinate
  descent.
\newblock {\em Journal of Statistical Software}, 33(1):1--22.

\bibitem[Gentleman and Geyer, 1994]{Gentleman.Geyer1994}
Gentleman, R. and Geyer, C.~J. (1994).
\newblock Maximum likelihood for interval censored data: Consistency and
  computation.
\newblock {\em Biometrika}, 81(3):618--623.

\bibitem[Guillaume et~al., 2017]{Guillaume.etal2017}
Guillaume, R., Couso, I., and Dubois, D. (2017).
\newblock Maximum likelihood with coarse data based on robust optimisation.
\newblock In {\em Proceedings of the {{Tenth International Symposium}} on
  {{Imprecise Probability}}: {{Theories}} and {{Applications}}}, pages
  169--180.

\bibitem[Heitjan, 1989]{Heitjan1989}
Heitjan, D.~F. (1989).
\newblock Inference from grouped continuous data: A review.
\newblock {\em Statistical Science}, 4(2):164--179.

\bibitem[Hjort and Pollard, 2011]{Hjort.Pollard2011}
Hjort, N.~L. and Pollard, D. (2011).
\newblock Asymptotics for minimisers of convex processes.

\bibitem[Huang, 1996]{Huang1996}
Huang, J. (1996).
\newblock Efficient estimation for the proportional hazards model with interval
  censoring.
\newblock {\em Annals of Statistics}, 24(2):540--568.

\bibitem[Kowal and Canale, 2020]{Kowal.Canale2020}
Kowal, D.~R. and Canale, A. (2020).
\newblock Simultaneous transformation and rounding ({{STAR}}) models for
  integer-valued data.
\newblock {\em Electronic Journal of Statistics}, 14(1):1744--1772.

\bibitem[Lee et~al., 2006]{Lee.etal2006}
Lee, S.-I., Lee, H., Abbeel, P., and Ng, A.~Y. (2006).
\newblock Efficient {{L1}} regularized logistic regression.
\newblock In {\em Aaai}, volume~6, pages 401--408.

\bibitem[Littmann et~al., 2022]{Littmann.etal2022}
Littmann, K., Hagstr{\"o}m, E., H{\"a}bel, H., Bottai, M., Eriksson, M.,
  Parini, P., and Brinck, J. (2022).
\newblock Plasma lipoprotein(a) measured in the routine clinical care is
  associated to atherosclerotic cardiovascular disease during a 14-year
  follow-up.
\newblock {\em European Journal of Preventive Cardiology}, 28(18):2038--2047.

\bibitem[Littmann et~al., 2019]{Littmann.etal2019}
Littmann, K., Wodaje, T., Alvarsson, M., Bottai, M., Eriksson, M., Parini, P.,
  and Brinck, J. (2019).
\newblock The {{Association}} of {{Lipoprotein}}(a) {{Plasma Levels With
  Prevalence}} of {{Cardiovascular Disease}} and {{Metabolic Control Status}}
  in {{Patients With Type}} 1 {{Diabetes}}.
\newblock {\em Diabetes Care}, 43(8):1851--1858.

\bibitem[McGough et~al., 2021]{McGough.etal2021}
McGough, S.~F., Incerti, D., Lyalina, S., Copping, R., Narasimhan, B., and
  Tibshirani, R. (2021).
\newblock Penalized regression for left-truncated and right-censored survival
  data.
\newblock {\em Statistics in Medicine}, 40(25):5487--5500.

\bibitem[Milzarek and Ulbrich, 2014]{Milzarek.Ulbrich2014}
Milzarek, A. and Ulbrich, M. (2014).
\newblock A semismooth {{Newton}} method with multidimensional filter
  globalization for {{L1-Optimization}}.
\newblock {\em SIAM Journal on Optimization}, 24:298--333.

\bibitem[Negahban et~al., 2009]{Negahban.etal2009}
Negahban, S., Yu, B., Wainwright, M.~J., and Ravikumar, P. (2009).
\newblock A unified framework for high-dimensional analysis of {{M-estimators}}
  with decomposable regularizers.
\newblock {\em Advances in neural information processing systems}, 22.

\bibitem[Negahban et~al., 2012]{Negahban.etal2012}
Negahban, S.~N., Ravikumar, P., Wainwright, M.~J., and Yu, B. (2012).
\newblock A unified framework for high-dimensional analysis of {{M-estimators}}
  with decomposable regularizers.
\newblock {\em Statistical Science}, 27(4).

\bibitem[Pr{\'e}kopa, 1973]{Prekopa1973}
Pr{\'e}kopa, A. (1973).
\newblock On logarithmic concave measures and functions.
\newblock {\em Acta Scientiarum Mathematicarum}, 34:335--343.

\bibitem[Ramsay, 1988]{Ramsay1988}
Ramsay, J.~O. (1988).
\newblock Monotone regression splines in action.
\newblock {\em Statistical Science}, 3(4):425--441.

\bibitem[Royston and Parmar, 2002]{Royston.Parmar2002}
Royston, P. and Parmar, M. K.~B. (2002).
\newblock Flexible parametric proportional-hazards and proportional-odds models
  for censored survival data, with application to prognostic modelling and
  estimation of treatment effects.
\newblock {\em Statistics in Medicine}, 21(15):2175--2197.

\bibitem[Taraldsen, 2011]{Taraldsen2011}
Taraldsen, G. (2011).
\newblock Analysis of rounded exponential data.
\newblock {\em Journal of Applied Statistics}, 38(5):977--986.

\bibitem[{The fsnet package}, 2022]{Thefsnetpackage2022}
{The fsnet package} (2022).
\newblock {{GitHub}}.
\newblock . https://github.com/koekvall/fsnet (accessed 09/09/2022).

\bibitem[Tutz and Schmid, 2016]{Tutz.Schmid2016}
Tutz, G. and Schmid, M. (2016).
\newblock {\em Modeling {{Discrete Time-to-Event Data}}}.
\newblock Springer {{Series}} in {{Statistics}}. {Springer International
  Publishing : Imprint: Springer}, {Cham}, 1st ed. 2016 edition.

\bibitem[{van de Vijver} et~al., 2002]{vandeVijver.etal2002}
{van de Vijver}, M.~J., He, Y.~D., {van't Veer}, L.~J., Dai, H., Hart, A.
  A.~M., Voskuil, D.~W., Schreiber, G.~J., Peterse, J.~L., Roberts, C., Marton,
  M.~J., Parrish, M., Atsma, D., Witteveen, A., Glas, A., Delahaye, L., {van
  der Velde}, T., Bartelink, H., Rodenhuis, S., Rutgers, E.~T., Friend, S.~H.,
  and Bernards, R. (2002).
\newblock A gene-expression signature as a predictor of survival in breast
  cancer.
\newblock {\em The New England Journal of Medicine}, 347(25):1999--2009.

\bibitem[Wang and Yan, 2021]{Wang.Yan2021}
Wang, W. and Yan, J. (2021).
\newblock Shape-restricted regression splines with {{R}} package splines2.
\newblock {\em Journal of Data Science}, 19(3):498--517.

\bibitem[Yuan et~al., 2012]{Yuan.etal2012}
Yuan, G.-X., Ho, C.-H., and Lin, C.-J. (2012).
\newblock An improved glmnet for l1-regularized logistic regression.
\newblock {\em Journal of Machine Learning Research}, 13(64):1999--2030.

\bibitem[Zeng et~al., 2016]{Zeng.etal2016}
Zeng, D., Mao, L., and Lin, D. (2016).
\newblock Maximum likelihood estimation for semiparametric transformation
  models with interval-censored data.
\newblock {\em Biometrika}, 103(2):253--271.

\end{thebibliography}

% \section*{Supporting information}
% Web Appendices and Figures referenced in Sections  \ref{sec:intro},
% \ref{sec:model}, \ref{sec:asy}, and \ref{sec:data_ex}, as well as code for
% the simulations and data examples, are available with this paper at the
% Biometrics website on Wiley Online Library. The code is also available at
% \url{https://github.com/koekvall/finite-suppl/}.
\end{document}